\documentclass[prd, onecolumn, amsmath,amssymb,superscriptaddress,nofootinbib,12pt]{revtex4-2}
\usepackage{url}

\usepackage{epsfig}
\usepackage{amsfonts}
\usepackage{graphicx}
\usepackage{epsfig}
\usepackage{eepic}
\usepackage{amsmath}
\usepackage{amssymb}
\usepackage{color}
\usepackage{bbm}
\usepackage{dcolumn}
\usepackage{bm}
\usepackage[normalem]{ulem}
\usepackage{mathrsfs}
\usepackage{bbold}
\usepackage{datetime}

\usepackage{overpic} 
\usepackage{rotating}
\usepackage[usenames,dvipsnames]{xcolor}
\usepackage[colorlinks=true,citecolor=Magenta,linkcolor=Green,urlcolor=Black]{hyperref}
\usepackage{lipsum} 
\usepackage{tikz,tikz-3dplot}
\usetikzlibrary{shapes.geometric}
\usepackage{etoolbox} 
\usepackage[capitalize]{cleveref}
\usepackage{extarrows}

\def\bc{\begin{center}}

\def\ec{\end{center}}
\def\be{\begin{eqnarray}}
\def\ee{\end{eqnarray}}
\definecolor{dyellow}{rgb}{1.,0.8,.0}
\definecolor{myblue}{rgb}{.1,.1,.7}
\definecolor{dcyan}{rgb}{.0,.6,.6}
\definecolor{dmagenta}{rgb}{0.6,0.0,0.6}
\definecolor{brown}{rgb}{0.6,0.2,0.}
\definecolor{darkblue}{rgb}{.0,.0,0.5}
\definecolor{darkred}{rgb}{0.75,0.0,0.0}
\definecolor{orange}{rgb}{1.,.6,.0}
\definecolor{dorange}{rgb}{0.8,.4,.0}
\definecolor{darkgreen}{rgb}{0.0,0.6,0.0}
\definecolor{purple}{rgb}{.4,.0,.4}
\definecolor{lightgrey}{rgb}{0.7, 0.7, 0.7}
\definecolor{grey}{rgb}{0.4, 0.4, 0.4}

\usepackage{geometry}
\usepackage[position=t, singlelinecheck=off]{subfig}
\usepackage[font=small,labelfont=bf,justification=raggedright]{caption}

\newcommand{\xdownarrow}[1]{%
  {\left\downarrow\vbox to #1{}\right.\kern-\nulldelimiterspace}
}
\newcommand{\xuparrow}[1]{%
  {\left\uparrow\vbox to #1{}\right.\kern-\nulldelimiterspace}
}

\definecolor{myred}{RGB}{189, 38, 49}

\begin{document}
\title{Learning topological defects formation with neural networks in a quantum phase transition}
\author{Han-Qing Shi} \email{by2030104@buaa.edu.cn}
\affiliation{Center for Gravitational Physics, Department of Space Science, Beihang University, Beijing 100191, China}
\author{Hai-Qing Zhang} \email{hqzhang@buaa.edu.cn}
\affiliation{Center for Gravitational Physics, Department of Space Science, Beihang University, Beijing 100191, China}
\affiliation{Peng Huanwu Collaborative Center for Research and Education, Beihang University, Beijing 100191, China}

\begin{abstract}
Neural networks possess formidable representational power, rendering them invaluable in solving complex quantum many-body systems. While they excel at analyzing static solutions, nonequilibrium processes, including critical dynamics during a quantum phase transition, pose a greater challenge for neural networks. To address this, we utilize neural networks and machine learning algorithms to investigate the time evolutions, universal statistics, and correlations of topological defects in a one-dimensional transverse-field quantum Ising model. Specifically, our analysis involves computing the energy of the system during a quantum phase transition following a linear quench of the transverse magnetic field strength. The excitation energies satisfy a power-law relation to the quench rate, indicating a proportional relationship between the excitation energy and the kink numbers. Moreover, we establish a universal power-law relationship between the first three cumulants of the kink numbers and the quench rate, indicating a binomial distribution of the kinks. Finally, the normalized kink-kink correlations are also investigated and it is found that the numerical values are consistent with the analytic formula.

\end{abstract}

\maketitle

{\bf Keywords:} Neural networks, Machine learning, Transverse-field quantum Ising model,

 \hspace{2.2cm} Kibble-Zurek mechanism


\newpage

\section{Introduction}
\label{intro}
One of the most challenging problems in modern physics is the so-called many-body problem. In its quantum version -- quantum many-body physics, the exponential complexity of the states in the Hilbert space makes the strongly correlated systems difficult to deal with \cite{fetter2012quantum}. Only limited analytical solutions are amenable to a few simple models \cite{albeverio2012solvable}. Therefore, resorting to a powerful algorithm by using suitable parameters to generalize the physical states becomes an intriguing direction. Along this way, many numerical methods are proposed, such as density matrix renormalization group \cite{white1992density} and quantum Monte Carlo \cite{ceperley1986quantum,troyer2005computational}.  
Although these methods work well for some specific problems, they lack universalities. 

 Fortunately, neural network methods have more universalities. The same neural network can be used to represent the states or to study the dynamical processes for various systems, such as those with different dimensions or with different interactions.
The recent state-of-the-art neural networks have been shown to provide high efficient representations of such complex states, making the overwhelming complexity computationally tractable \cite{carleo2019machine,karagiorgi2022machine}. Except for the success in the industrial applications, such as the image and speech recognitions \cite{lecun2015deep}, the autonomous driving, and the game of Go \cite{silver2016mastering}, neural networks have been widely adopted to study a broad spectrum of areas in physics, ranging from statistical and quantum physics to high energy and cosmology \cite{lam2021machine,raissi2019physics,carrasquilla2017machine, shi2019neural, park2020geometry}.   
    
  Among these successful applications in physical sciences, the more challenging task is to use neural networks to study nonequilibrium problems. Recently, an algorithm of artificial neural networks was proposed to solve the unitary time evolutions in a quantum many-body system \cite{carleo2017solving}. Later developments in this direction can be found in \cite{schmitt2020quantum,hartmann2019neural,Hofmann:2021bbi,schmitt2018quantum,Schmitt:2021jca,czischek2018quenches,fabiani2019investigating,luo2022autoregressive,reh2021time,donatella2022dynamics,gutierrez2022real}. 
In nonequilibrium dynamics, a common issue is the emergence of universalities during the critical dynamics of a phase transition. One of the most well-known challenges in this regard is the formation of topological defects. It is stated that topological defects will arise in the course of a phase transition with symmetry breaking of the system due to the celebrated Kibble-Zurek mechanism (KZM) \cite{kibble1976topology,zurek1985cosmological}. Number density of topological defects was found to satisfy a universal power-law with respect to the quench rate.  
The formation of topological defects and the KZM have been widely examined in various numerical simulations and experiments, including in quantum phase transition \cite{zurek2005dynamics,del2018universal}, in quantum field theory with matrix product states \cite{gillman2018kibble}, in AdS/CFT correspondence \cite{Sonner:2014tca,Chesler:2014gya,Zeng:2019yhi,Li:2021jqk}, in programmable quantum simulators \cite{keesling2019quantum,ebadi2021quantum} and in D-wave devices \cite{weinberg2020scaling}, to name some relevant references.  

\begin{figure}[t]
\begin{center}
\includegraphics[trim=0.cm 0.cm 0cm 0.cm, clip=true, scale=0.6]{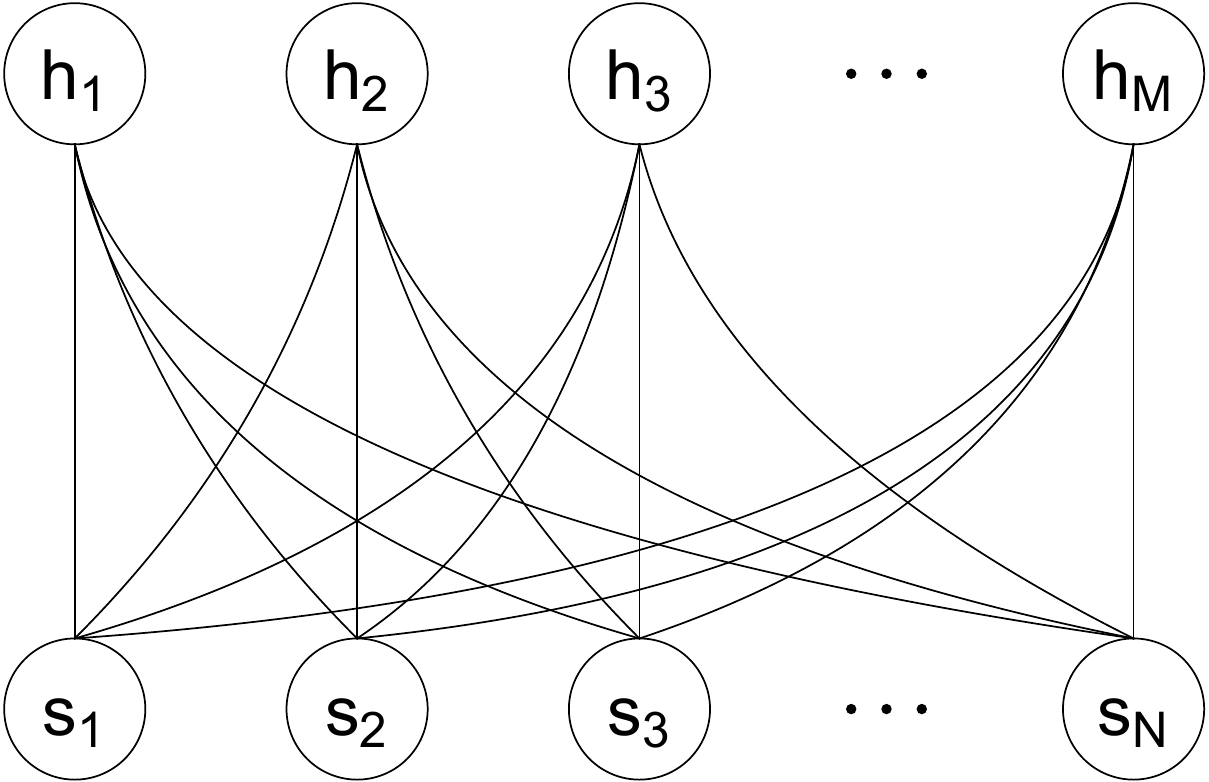}
\caption{Sketchy map of the structure of the RBM. RBM has a hidden layer with $h_i$ $(i=1,\cdots M)$ as hidden neurons  and a visible layer with $s_j$ $(j=1,\cdots N)$ as visible neurons. The lines linking the hidden points and visible points represent the interactions. There is no intralayer interactions in the hidden layer or visible layer themselves.}
\label{rbmpic}
\end{center}
\end{figure}

In \cite{carleo2017solving} the machine learning methods was merely applied in the unitary dynamics without phase transitions. Critical dynamics, i.e., the dynamics across the critical point of a phase transition is more complex and has richer phenomena \cite{sachdev2011quantum}. Critical slowing down near the phase transition point may invalidate the applicability of this method. In this paper, we extend the machine learning methods introduced in \cite{carleo2017solving} to study the nonequilibrium process of critical dynamics in a one-dimensional transverse-field quantum Ising model (TFQIM). TFQIM is a widely used model to study the phase transitions of one-dimensional spin chain and has been extensively studied analytically or experimentally such as in \cite{dziarmaga2005dynamics,nowak2021quantum,King:2022phl}. Therefore, TFQIM is a very suitable testbed to check the representing accuracy of neural networks and the robustness of machine learning methods.  Specifically, we study the time evolutions of the energy, universal statistics and correlations of the topological defects formed in the TFQIM after quantum phase transition induced by a quench. In particular, we quench the strength of the transverse magnetic field to drive the system from a paramagnetic state into a ferromagnetic state, during which the topological defects, i.e. the kinks where the polarization of the spins changes their directions, will form due to the KZM. In the machine learning we introduce the Restricted Boltzmann Machine (RBM) as a representation of the quantum state for TFQIM.  RBM is a kind of neural networks with two layers of neurons, i.e. visible layer and hidden layer (see Fig.\ref{rbmpic}). In order to solve the ground state and the time evolution of the system, the stochastic reconfiguration (SR) method and time-dependent variational Monte Carlo (VMC) approach \cite{sorella2001generalized} are utilized, respectively.  We find that time evolutions of the energy  expectation value from the neural networks are perfectly consistent with the results reported in \cite{dziarmaga2005dynamics}. After the quench, the excited energy of the system are found to satisfy a power-law relation against the quench rate, which reveals the proportional relationship between the excitation energy and the kink numbers. Besides, the counting statistics of the kink numbers satisfy the Poisson binomial distributions introduced previously in \cite{del2018universal,gomez2020full}. By computing the first three cumulants of the kink pair numbers, we find that they satisfy a universal power-law scalings to the quench rate consistent with the theoretical predictions.  Additionally, we compute the kink-kink correlations at the end of the quench. The numerical data match the analytic formula presented in \cite{nowak2021quantum} very well. Therefore, our results show a very high accuracy of neural networks to investigate the critical dynamics of TFQIM. 

\section{Results}
\label{sec2}
\subsection{Quantum quench of TFQIM }
 We study the formation of topological defects in one-dimensional TFQIM, with the Hamiltonian of a spin chain of $N$ sites in a transverse magnetic field \cite{dziarmaga2005dynamics},
\begin{equation}\label{H}
H=-J\sum_{i=1}^N\left(\sigma_i^z\sigma_{i+1}^z+h\sigma_i^x\right),
\end{equation}
where $\sigma^{z}_i$ and $\sigma^x_i$ are the Pauli matrices at the site $i$ in the $z$ and $x$ directions, while $J$ and $h$ respectively denote the coupling strengths between the nearest-neighbor sites and transverse magnetic field strength. We consider the periodic boundary conditions (PBC) for this spin chain by imposing $\vec\sigma_{N+1}=\vec\sigma_1$ with even $N$ for simplicity. There exists a quantum phase transition at the critical point $|h_c|=1$. Without loss of generality, we will only focus on the regime $h\geq0$.  As $h\gg1$, the ground state is in a paramagnetic state; On the other hand, while $h\ll1$ the ground state is in the two-degenerated ferromagnetic states with spins up or down along the $z$-direction. 
In the limit of $N\to\infty$ the energy gap at the critical point $h_c=1$ tends to zero, which resembles the critical slowing down. Therefore, if the system goes from the regime $h>1$ to the regime $h<1$, it is impossible to cross the critical point without exciting the system. As a result, the system will end up in the configurations that spins will point up or point down in some finite domains. Consequently, the kinks, a kind of topological defects in one dimension, form.

Conventionally, one evolves the system by linearly quenching the transverse magnetic field as,
\begin{equation}\label{quench}
h(t)=-\frac{t}{\tau_Q},\qquad t\in[-T,0]
\end{equation} 
in which $\tau_Q$ is the quench rate. In the initial time $T/\tau_Q\gg 1$, we prepare the ground state in a strong transverse magnetic field, thus the spins will all point up along the transverse field direction.  The system will evolve according to the quench profile \eqref{quench}, cross the critical point and then end at $h(t=0)=0$.  During this phase transition, kinks for the polarization directions of spins will form according to KZM. The analytic solutions to the dynamics of the Hamiltonian \eqref{H} were previously carried out in \cite{dziarmaga2005dynamics}. In the limit of $N\to\infty$ , the average number density of kinks $\langle\mathcal{\hat N}\rangle$  (where $\hat{\mathcal{N}}=\frac{1}{2}\sum_{i=1}^N(1-\sigma^z_i\sigma^z_{i+1})$ is the kink number operator) are found to satisfy a universal power-law to the quench rate as $\langle \mathcal{\hat N}\rangle\propto \tau_Q^{-1/2}$, consistent with the KZM.


 \begin{figure}[h]
\begin{center}
	\includegraphics[trim=0.cm -7mm 0cm 0.cm, clip=true, scale=0.55]{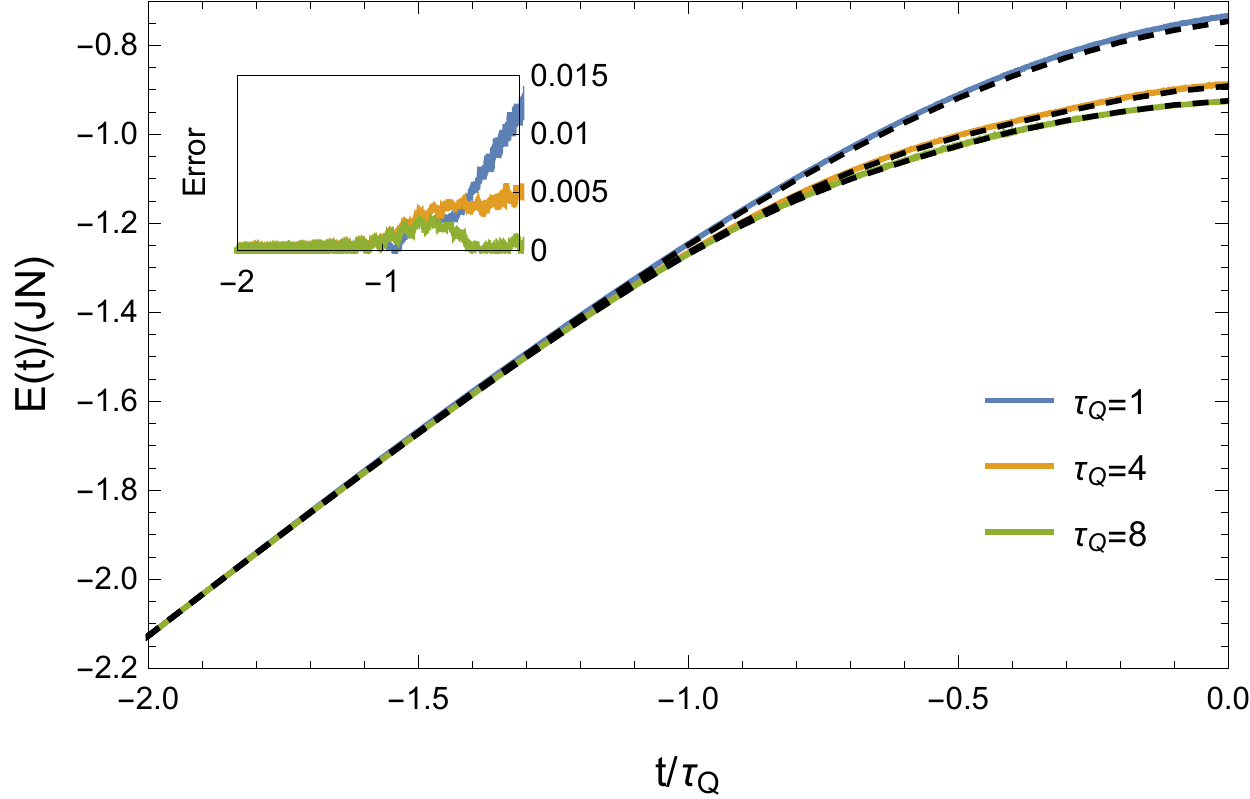}
	\includegraphics[trim=3.cm 9.cm 3cm 10.cm, clip=true, scale=0.46]{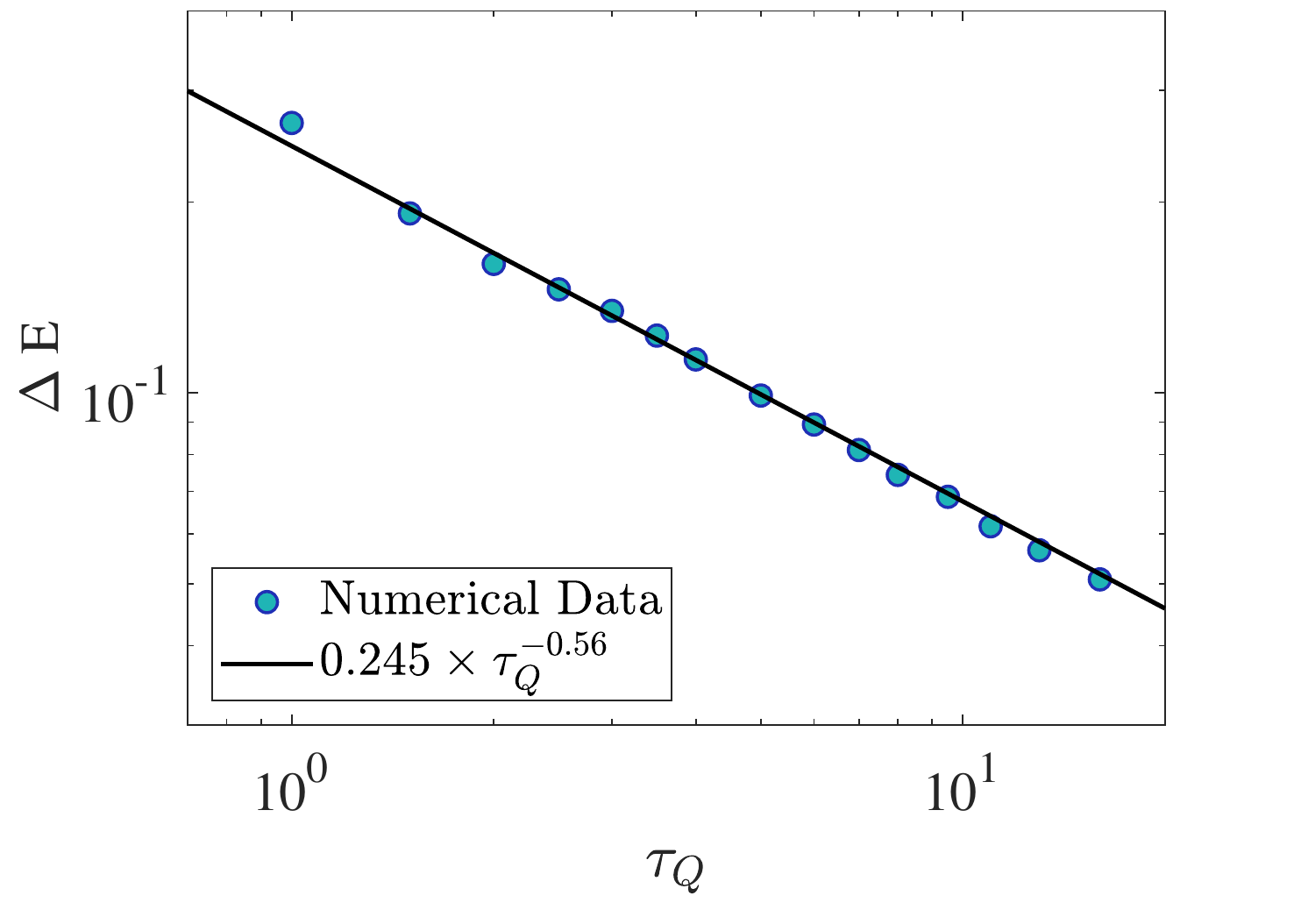}
\caption{{(Left)Time evolution of the energy expectation value $E(t)/J$ per lattice point with respect to the reduced time $t/\tau_Q$.} The solid lines represent the evolution of energy for three different quench rates from machine learning methods,  while the black dashed lines are the comparing results from the methods in \cite{dziarmaga2005dynamics}. The inset plot exhibits the relative errors for both methods, from which we see that they match each other very well. (Right) The excitation energy density $\Delta E$ with respect to $\tau_Q$ at the end of the quench. The fitting line has a power-law scaling as $\Delta E\approx 0.245\times\tau_Q^{-0.56}$, which indicates that the excitation energy is proportional to the kink numbers. }
\label{energypic}
\end{center}
\end{figure}

\subsection{Time evolution of energy expectation value}
 Utilizing the machine learning method (see Section \ref{methods} ``Materials and Methods"), we consider the time evolution of a one-dimensional TFQIM \eqref{H} under a linear quench \eqref{quench} through the critical point with various quench rates. We set PBC to the spin chains, thus they satisfy the even parity \cite{dziarmaga2005dynamics}. The coupling strengths between the nearest-neighbor site $J$ and the lattice spacing were set to be unit. The number of lattice points we take is $N=100$ and the time period is $t\in[-2\tau_Q,0]$, corresponding to the strength of transverse magnetic field $h$ from 2 to 0. The hidden sites we take is $M=4N$. The time step we use is $\Delta t=1/1000$. Then the system will evolve under the time-dependent VMC method. In the left plot of Fig.\ref{energypic}, we show the evolution of energy expectation value for three different quench rates $\tau_Q=1,4 \ {\rm and}\  8$. We can see that the solutions from the machine learning methods (solid lines) and the solutions from the analytic methods in \cite{dziarmaga2005dynamics} (black dashed lines) match each other very well. In particular, we find that in the early time the energies satisfy a linear function as $E(t)/(JN)\approx t/\tau_Q$, which can be readily derived from the Hamiltonian \eqref{H} and the quench profile \eqref{quench}, since in the early time the term $h\sigma^x_i$ makes a dominant contribution to the Hamiltonian. Then, roughly in the intermediate time $t/\tau_Q\in[-1,-0.5]$,  faster quench (smaller $\tau_Q$) increases the energy faster, leading to a higher final energy as the quench ends ($t/\tau_Q=0$).  The relative errors for both methods at the end of the quench are relatively small, specifically the errors are $(1.41\%, 0.51\%, 0.02\%)$  for $\tau_Q=(1,4,8)$, respectively, as the inset plot shows in the left plot of Fig.\ref{energypic}.

 Since each kink has the same energy at $h=0$, we can determine the number of kinks by analyzing their excitation energy.  In the right plot of Fig.\ref{energypic}, we show the excitation energy density $\Delta E$ at $t=0$ with respect to the quench rate $\tau_Q$. The excitation energy density is defined as
\be \Delta E=(E-E_{0})/(JN)\ee
where $E_0$ is the ground state energy as $h=0$ in the Eq.\eqref{H}. The fitting line is roughly $\Delta E\approx0.245\times\tau_Q^{-0.56}$. We see that the power-law $\tau_Q^{-0.56}$ is close to the power-law predicted by KZM that the mean kink number density is proportional to the quench rate as $\langle\mathcal{\hat N}\rangle\propto\tau_Q^{-0.5}$ (We will investigate it in the following in detail.). This consistency demonstrates our assumption that the excitation energy is proportional to the kink numbers.

\subsection{Statistics of kink numbers beyond KZM}
The topological defects, i.e., kinks form in the course of the quantum phase transitions due to the KZM.  It predicts that the power-law scalings of the mean value of the kink numbers to the quench rate is proportional to $\tau_Q^{-d\nu/(1+\nu z)}$, where $d=1$ for the one-dimensional spin chain; $\nu$ and $z$ are respectively the static and dynamic critical exponents. For the TFQIM they are $\nu=z=1$. Therefore, the theoretical prediction of the power law scaling between the mean kink number and the quench rate is $\langle\mathcal{\hat N}\rangle\propto\tau_Q^{-1/2}$.
However, the KZM predicts only the power-law relation between the average number of the kinks and the quench rate.  From \cite{del2018universal} the distribution of kinks in the TFQIM is assumed to satisfy the Poisson binomial distributions.  Consequently, the fluctuations away from the mean of kink numbers, such as the variance, satisfy a universal power-law scaling to the quench rate as well. Therefore, it will be vital to study these universal power-laws beyond KZM by virtue of the neural networks.  
These universal power-laws beyond KZM can be achieved by computing the higher order cumulants of the kink numbers. Since we adopt the PBC in the spin chain, the outcomes of the kink numbers are all even \cite{cui2020experimentally,bando2020probing}. Thus, in practice we compute the higher cumulants for the numbers of kink pairs, i.e., $\mathcal {\hat N}_P=\mathcal{\hat N}/2$.  According to \cite{del2018universal,gomez2020full}, all the cumulants should be proportional to the mean, i.e.,  $\kappa_q\propto \langle\mathcal {\hat N}_P\rangle$  where $q$ are positive integers.  Due to the cost of time and the limited resources of the computer, we only compute the first three cumulants, i.e.  $\kappa_q$ with $q=(1,2,3)$ of the kink pair numbers. \footnote{ For the sites number $N=100$ and the neural network size $\alpha=M/N=4$, it would take approximately 48 hours to run the code once with $\tau_Q=16$.  If increasing the size of $N$ or $\alpha$, or bigger $\tau_Q$, the consumption of time will be much expensive. Therefore, considering the accuracy of the results and the cost of time, we set $N=100$, $\alpha=4$ and the greatest quench rate is $\tau_Q=16$. } Specifically, these cumulants can be expressed as  
\begin{eqnarray}\label{cumulants}
\kappa_1=&\langle\hat{\mathcal{N}}_P\rangle, \\
\kappa_2=&\langle\hat{\mathcal{N}}_P^2\rangle-\langle\hat{\mathcal{N}}_P\rangle^2, \\
\kappa_3=&\langle(\hat{\mathcal{N}}_P-\langle\hat{\mathcal{N}}_P\rangle)^3\rangle.
\end{eqnarray}
In other words, $\kappa_1$ is the mean value of the kink pair numbers; $\kappa_2$ is the variance of the kink pair numbers while $\kappa_3$ is related to the skewness of the kink pair numbers as $\kappa_3={\rm Skew}(\mathcal{\hat N}_P)\kappa_2^{3/2}$. 

\begin{figure}[t]
\begin{center}
\includegraphics[trim=3.cm 9.3cm 4cm 10cm, clip=true, scale=0.6]{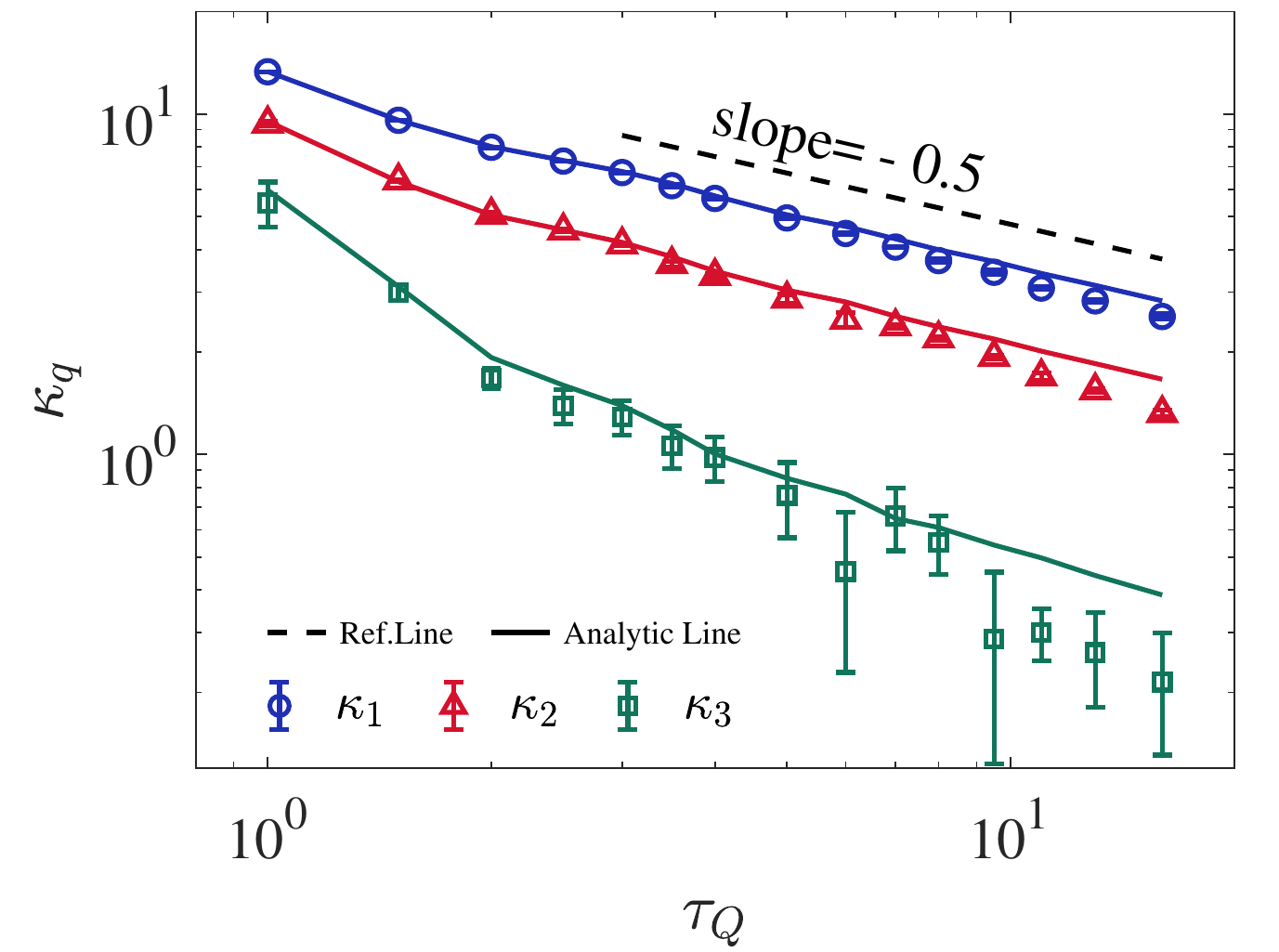}
\caption{ Double logarithmic plots for the cumulants $\kappa_q$ ($q=1,2,3$) of the kink pair number distributions with respect to quench rate $\tau_Q$. The circles, triangles and squares are the data from the neural network methods, while the solid lines are from the analytic method introduced in \cite{dziarmaga2005dynamics}.  The dashed line is the reference line with theoretical power law $\tau_Q^{-0.5}$. The error bars represent the standard errors. }
\label{kinkpic}
\end{center}
\end{figure}

In Fig.\ref{kinkpic}, we show the first three cumulants of the kink pair number distribution $\kappa_q,(q=1,2,3)$ as a function of the quench rate $\tau_Q$. The reference line (dashed line) represents the theoretical scaling $\kappa_q\propto\tau_Q^{-0.5}$.  The numerical data from the neural networks methods are shown in the circles ($\kappa_1$), triangles  ($\kappa_2$) and squares ($\kappa_3$). The error bars indicate the standard errors in the statistics. The solid lines are obtained from the analytic methods used in \cite{dziarmaga2005dynamics}. (In order not to make confusions, we stress that the solid lines are {\it not} the fitting lines for the numerical data from the neural networks method.) By fitting these cumulants, the power law scaling $\tau_Q^{\beta}$ for $\kappa_{1,2,3}$ have the powers $\beta=(-0.58,-0.61,-0.64)$, respectively.  The fitting range are taken from $\tau_Q\geq3$ for $\kappa_{1,2}$ and $3\leq\tau_Q\leq8$ for $\kappa_3$. For fast quench (relatively small quench rates), i.e. $\tau_Q\lesssim3$, the scaling relations will deviate from the KZM power-laws because of the finite size effect \cite{del2018universal,gomez2020full}.

 
From the Fig.\ref{kinkpic}, we can see that the results of the power $\beta$, in particular for $\kappa_1$ and $\kappa_2$, is a little bit away from the theoretical power $-1/2$ since we are using the finite size $N=100$. In the limit of large $N$, this power-law would tend to the theoretical predictions as \cite{del2018universal,dziarmaga2005dynamics,gomez2020full} demonstrated. Neverthelss, the power $\beta_{\kappa_1}=-0.58$ for $N=100$ in our case is perfectly consistent with the numerical results in \cite{zurek2005dynamics} where the authors also adopted $N=100$ sites. It should be noted that $\kappa_3$ deviates relatively large compared to the first two cumulants. This is because $\kappa_3$ is more sensitive to the sites number $N$. Similar results in experiments, numerical simulations and holography were reported previously in \cite{gomez2020full,cui2020experimentally,bando2020probing,delCampo:2021rak}. In the Appendix \ref{sec:a}, we compare the numerical results for the cumulants $\kappa_{1,2,3}$ with different $N$'s. And we find that as $N$ increases, the power-law behavior will improve and get closer to the theoretical predictions. 

Moreover, from Fig.\ref{kinkpic} we notice that the value of the ratio $\kappa_2/\kappa_1$ is roughly $\kappa_2/\kappa_1\approx 0.30$ from the fitting, which is very close to the theoretical predictions $\kappa_2/\kappa_1=(2-\sqrt2)/2\approx 0.29$ in \cite{del2018universal,gomez2020full}. Besides, the ratio $\kappa_3/\kappa_1\approx 0.042$ in our case is a little bit away from the predictions $\kappa_3/\kappa_1=(1-3/\sqrt2+2/\sqrt3)\approx 0.033$ in \cite{del2018universal,gomez2020full}. The reason is that $\kappa_3$ is more sensitive to the sites number $N$ as we stated above. It is expected that increasing the sites number would improve this problem.

\subsection{Kink-kink correlations}

\begin{figure}[t]
	\begin{center}
		\includegraphics[trim=3.cm 9.cm 3cm 9.5cm, clip=true, scale=0.6]{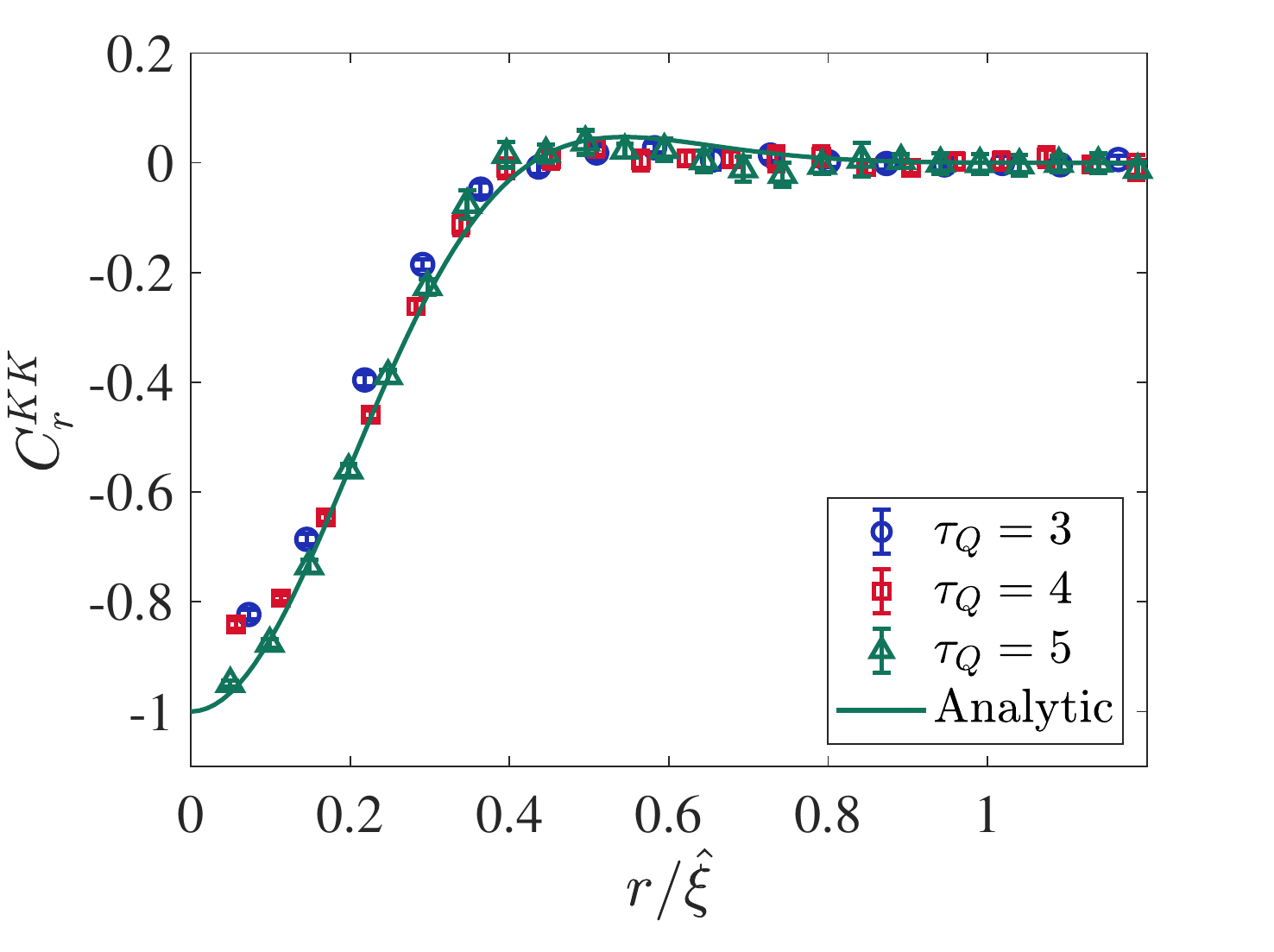}
		\caption{The normalized kink-kink correlations $C_r^{KK}$ against the normalized distances $r/\hat\xi$. The green line is from the analytic formula in Eq.\eqref{analytic} with $\tau_Q=5$. The error bars in the numerical data represent the standard error.} 
		\label{c2}
	\end{center}
\end{figure}

Correlations between kinks are the delicate quantum features for the Ising model \cite{nowak2021quantum}, and it has been successfully tested on a one-dimensional transverse field Ising chain using programmable quantum annealer \cite{King:2022phl}. Therefore, it will be very important to check the accuracy of the neural network methods by studying the kink-kink correlation $C_r^{KK}$.  Following \cite{King:2022phl}, we define the normalized connected kink-kink correlation as 
\begin{equation}
C_r^{KK}=\frac1N\sum_{i=1}^N\frac{\left(\langle K_iK_{i+r}\rangle-\bar n^2\right)}{\bar n^2},
\end{equation}
where $K_i=\frac12(1-\sigma^z_i\sigma^z_{i+1})$ is the kink number operator at the site $i$, $\bar n=\langle\mathcal{\hat N}\rangle/N$ is the average kink number density.  The analytic formula for the normalized kink-kink correlation can be read out from \cite{nowak2021quantum} as,
\be\label{analytic}
C_r^{KK}=\alpha \frac{\hat \xi}{l}\left(\frac rl\right)^2e^{-3\pi(r/l)^2}-e^{-2\pi(r/\hat\xi)^2}.
\ee
in which $\alpha=\frac{9747\pi}{3200}\approx9.57$ is a numerical factor, $\hat\xi$ is the typical KZ correlation length that indicates the average distance between kinks, i.e., $\hat \xi= N/\langle\mathcal{\hat N}\rangle=1/\bar n$ and $l=\hat\xi\sqrt{1+\left(3\log\tau_Q/(4\pi)\right)^2}$ is the correlation range. In Fig.\ref{c2}, we show the numerical results of $C_r^{KK}$ against the normalized distance $r/\hat \xi$ for various quenches $\tau_Q=3,4$ and $5$. The green line is from the analytic formula Eq.\eqref{analytic} with $\tau_Q=5$. ( For $\tau_Q=3, 4$ and $5$, the differences between the analytic formula Eq.\eqref{analytic} are tiny. In order to distinguish the figure clearly, we only plot the analytic lines for $\tau_Q=5$.) We see that the numerical data will collapse together and match the analytic line very well. In particular, as $r/\hat\xi\to0$ the correlation goes to $-1$ and there is a peak at around $r/\hat\xi\approx0.5$, which were already demonstrated in \cite{nowak2021quantum,King:2022phl}. The correlation tends to zero as the distance is very large, which satisfies the physical intuitions.  Therefore, we see that the neural network methods can also uncover the delicate quantum properties of the Ising model. 

\section{Discussions}
We have realized the time evolutions of the energy expectation value, the universal statistics of the topological defects numbers and the kink-kink correlations in a quantum phase transition of a TFQIM by virtue of the neural networks. The results were found to satisfy theoretical predictions. Thus, it numerically verifies that the neural network methods not only can extend but also work well for the critical dynamics in a quantum phase transition. 

In our paper, we used the network size $\alpha=M/N=4$. However, we find that even with smaller $\alpha$, the numerical results will still be consistent with theoretical predictions. In the Appendix \ref{sec:b}, we compare the effects of network size $\alpha=2$ and $\alpha=4$ on the kink number cumulants and the energy expectation values. We see that for $\alpha=2$ the numerical results are in agreement with the those for $\alpha=4$. All of these suggested that the neural networks method perform very well and showed high accuracy for the quantum phase transition. We expect that the neural networks and machine learning methods may shed light on further complex dynamics in quantum many-body physics. 


\section{Materials and Methods}
\label{methods}
\subsection{Machine learning with neural networks}
 The states of quantum system can be characterized by wave functions. One can approximate the wave functions with neural-network quantum states (NQS) as $\Psi_{\!N\!N}(s,\mathcal{W})$, where $s=(s_1,s_2\dots,s_N)$ denotes the spin configurations at each site directed in the $z$-direction basis, and $\mathcal{W}$ is the neural network parameter. Different $\mathcal{W}$'s correspond to different quantum states of the system. Therefore, suitable values of $\mathcal W$ can make the NQS describe the states with topological defects after the phase transition from paramagnetic state to ferromagnetic state. Our goal is find out this parameter $\mathcal W$ to simulate the process of phase transition and quantify the statistics and correlations of the kinks. In this work, we adopt the RBM as neural networks which work well for the quantum spin chain \cite{carleo2017solving}. It consists of two layers, one is the visible layers with $N$ visible neurons $s_j$ and the other is the hidden layers with $M$ hidden neurons $h_i$, refer to Fig.\ref{rbmpic}. There are interactions between hidden and visible layers, but no intralayer interactions between themselves.  According to the structure of the neural networks, the quantum state can be described as
\begin{eqnarray}
\Psi_{\!N\!N}(s,\mathcal{W})\!=
\sum_{\{h\}}\exp\!\left[\sum_ja_js_j+\sum_ib_ih_i+\sum_{i,j}w_{ij}h_is_j\right],
\end{eqnarray}
where $s=\{s_j\}$ denotes the spin configurations and $\mathcal{W}=\{a,b,w\}$ are the neural network parameters. When we set $\mathcal{W}$ to be complex, the wave function $\Psi_{\!N\!N}$ can represent the amplitudes and phases of the states. The parameter $h_i=\{-1,1\}$ are the hidden variables. Since there is no intralayer interactions in the hidden layers, the hidden variables can be traced out explicitly in the first step. Therefore, the wave function becomes
\begin{equation}
\Psi_{\!N\!N}(s,\mathcal{W})=e^{\sum_ja_js_j}\prod_{i=1}^M2\cosh[b_i+\sum_jw_{ij}s_j].
\end{equation}

In the machine learning, the number of independent parameters contributes to the computational complexity of the wave functions $\Psi_{\!N\!N}$. Fortunately, we can reduce the number of parameters from the symmetry of the system or some physical constraints.  When a system has a symmetry under some operation $\hat{T}$, we can impose a constraint for the neural network $\Psi_{\!N\!N}(s,\mathcal{W})=\Psi_{\!N\!N}(\hat{T}s,\mathcal{W})$. This constraint will largely reduce the number of independent network parameters and facilitate the computations. In our case, lattice translational symmetry is used to reduce the parameters from the order of magnitude $\mathcal{O}(M\times N)$ to $\mathcal{O}(M)$.
In this  paper, we need to evolve the system from an initial ground state with strong transverse magnetic field to a state without any transverse magnetic field. In order to obtain this initial state, we have to train a beginning wave function whose parameters are random complex numbers with machine learning algorithms, in which this learning is realized through minimizing the expectation value of the energy $\langle E\rangle=\langle\Psi_{\!N\!N}|H|\Psi_{\!N\!N}\rangle/\langle\Psi_{\!N\!N}|\Psi_{\!N\!N}\rangle$ with SR method. 

After preparing the initial ground state, the system will evolve according to the quench profile \eqref{quench}. Therefore, the wave function should also depend on time. To this end, we render the neural network parameters to be functions of time. Then, the parameters will be computed at every time step with the time-dependent VMC method, by minimizing the distances $\delta$ between the exact time evolution and the approximate variational evolution
\begin{eqnarray}
\delta={\rm dist}\left[\partial_t\Psi_{\!N\!N}\left(\mathcal{W}(t)\right), -iH\Psi_{\!N\!N}\left(\mathcal{W}(t)\right)\right],
\end{eqnarray}
where,
\begin{eqnarray}
{\rm dist}\left[\Psi',\Psi\right]\equiv{\rm \arccos}\sqrt{\frac{\langle\Psi'|\Psi\rangle\langle\Psi|\Psi'\rangle}{\langle\Psi'|\Psi'\rangle\langle\Psi|\Psi\rangle}}.
\end{eqnarray}
 For more detailed information, we refer readers to \cite{carleo2017solving}.
In order to decrease the influence of the noise from the Monte Carlo method, some regularization methods are needed. We utilize the singular-value decomposition (SVD) regularization method introduced in \cite{Hofmann:2021bbi}.  This method can eliminate the effect from Monte Carlo sampling noise and make the program more stable. More specifically, we decompose the matrix with SVD, and remove the singular value less than a tolerence $\lambda\sim 10^{-8}$. As the approximate wave function is obtained, the average number of topological defects $\langle \Psi_{\!N\!N}|\mathcal{\hat N}|\Psi_{\!N\!N}\rangle$ can be computed through the kink number operator $\hat{\mathcal{N}}$. 
 

\section{ACKNOWLEDGEMENTS}
We appreciate the illuminating discussions with Dr. Peng-Zhang He. 
This work was partially supported by the National Natural Science Foundation of China (Grants No.11875095 and 12175008). 

{\bf Conflict of interest.} The authors declare that they have no conflict
of interest.

\bibliography{ref1.bib}

\appendix

\section{Finite-size Effects on the Cumulants}
\label{sec:a}
 In  Ref.\cite{dziarmaga2005dynamics}, it has been analytically confirmed that as the sites number $N$ goes to infinity, the scaling of the mean kink number to the quench rate will be proportional to $\tau_Q^{-0.5}$ which satisfies the KZM's prediction. Later this scaling law was extended to higher cumulants of the kink numbers \cite{del2018universal,gomez2020full}, and it was found that all cumulants should be parallel to each other in the large $N$ limit.
 
\begin{figure}[t]
	\begin{center}
	\includegraphics[trim=3cm 9cm 4.4cm 9cm, clip=true, scale=0.5, angle=0]{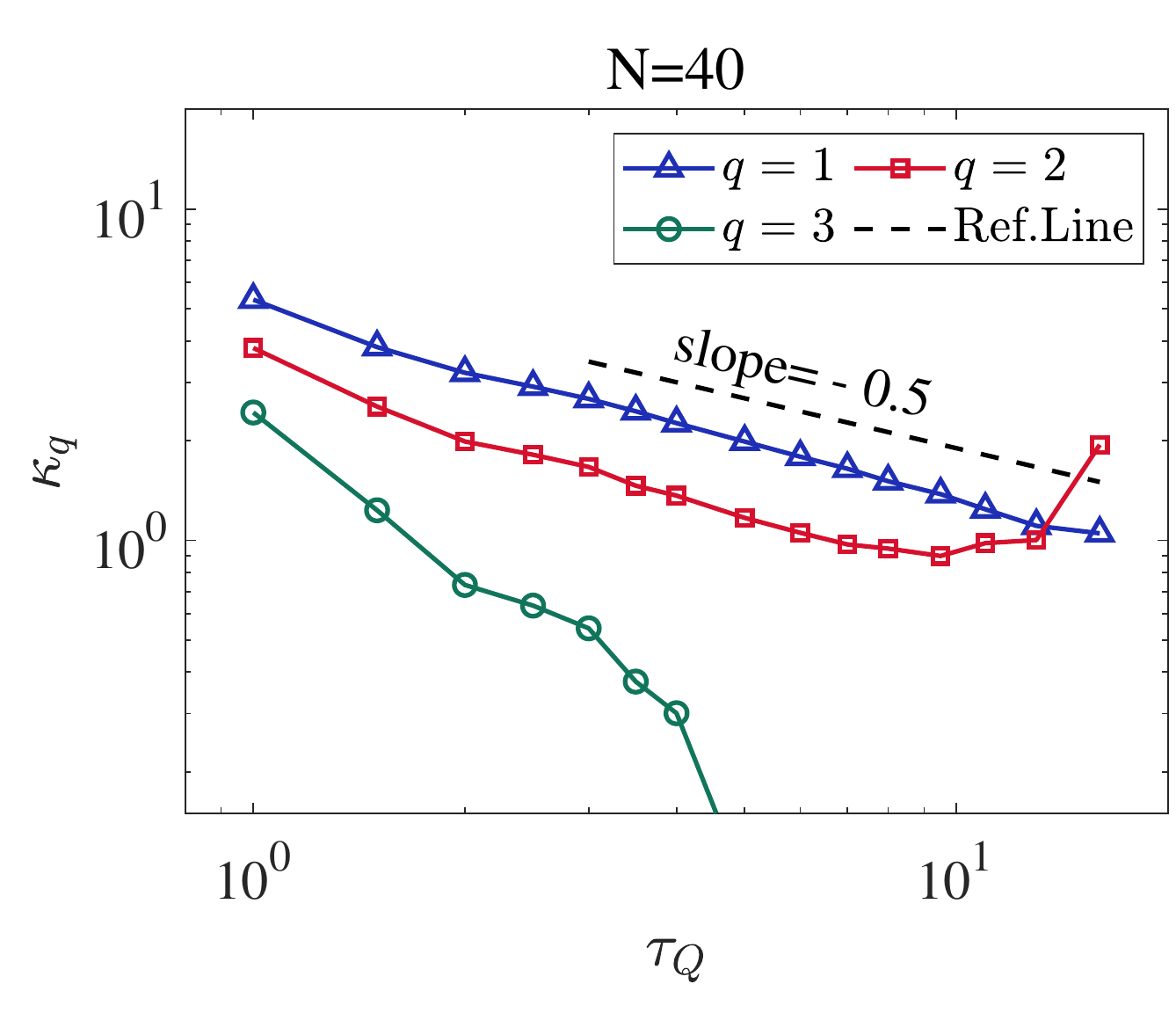}
	\includegraphics[trim=3cm 9cm 4.4cm 9cm, clip=true, scale=0.5, angle=0]{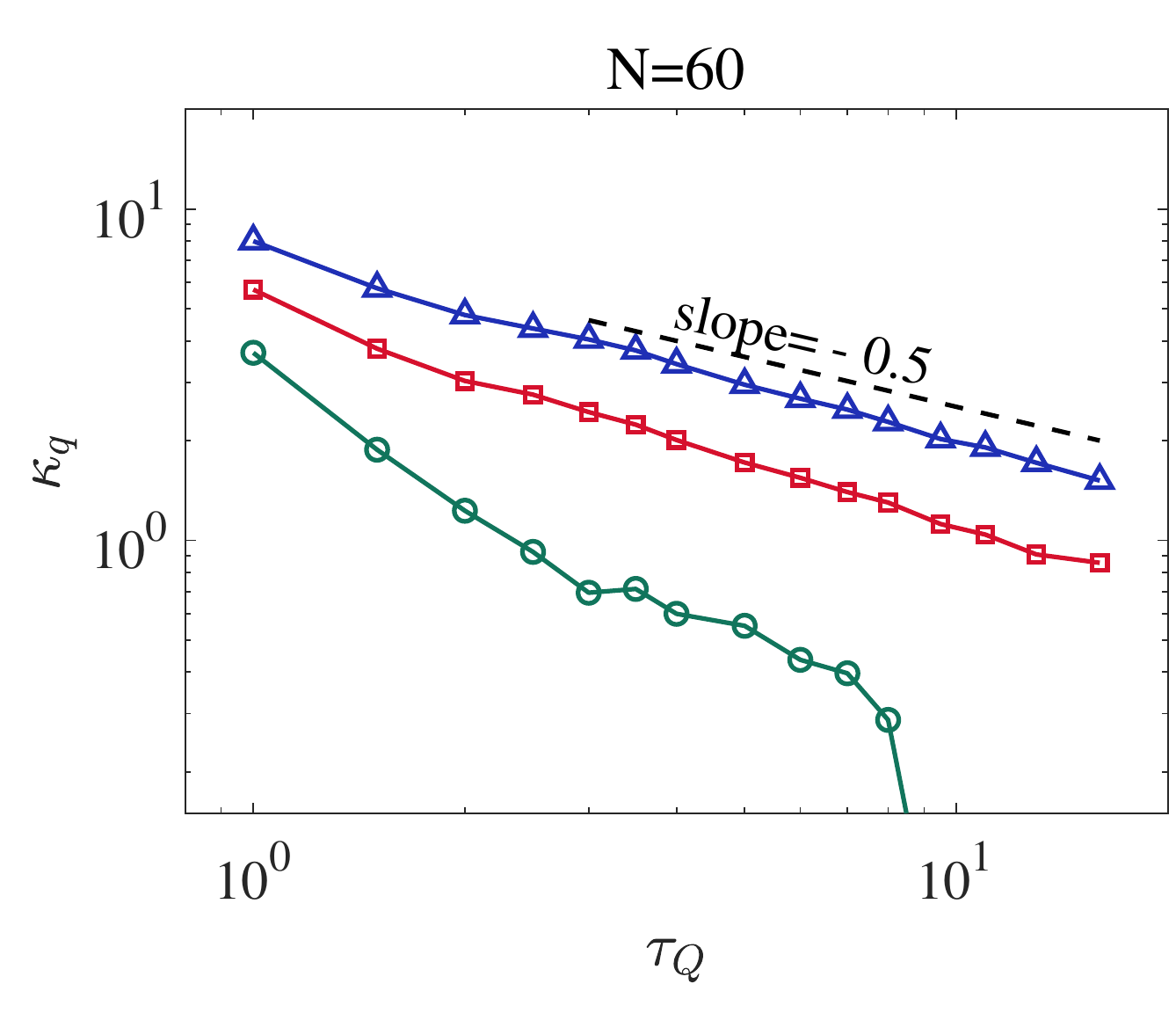}\\
	\includegraphics[trim=3cm 9cm 4.4cm 9cm, clip=true, scale=0.5, angle=0]{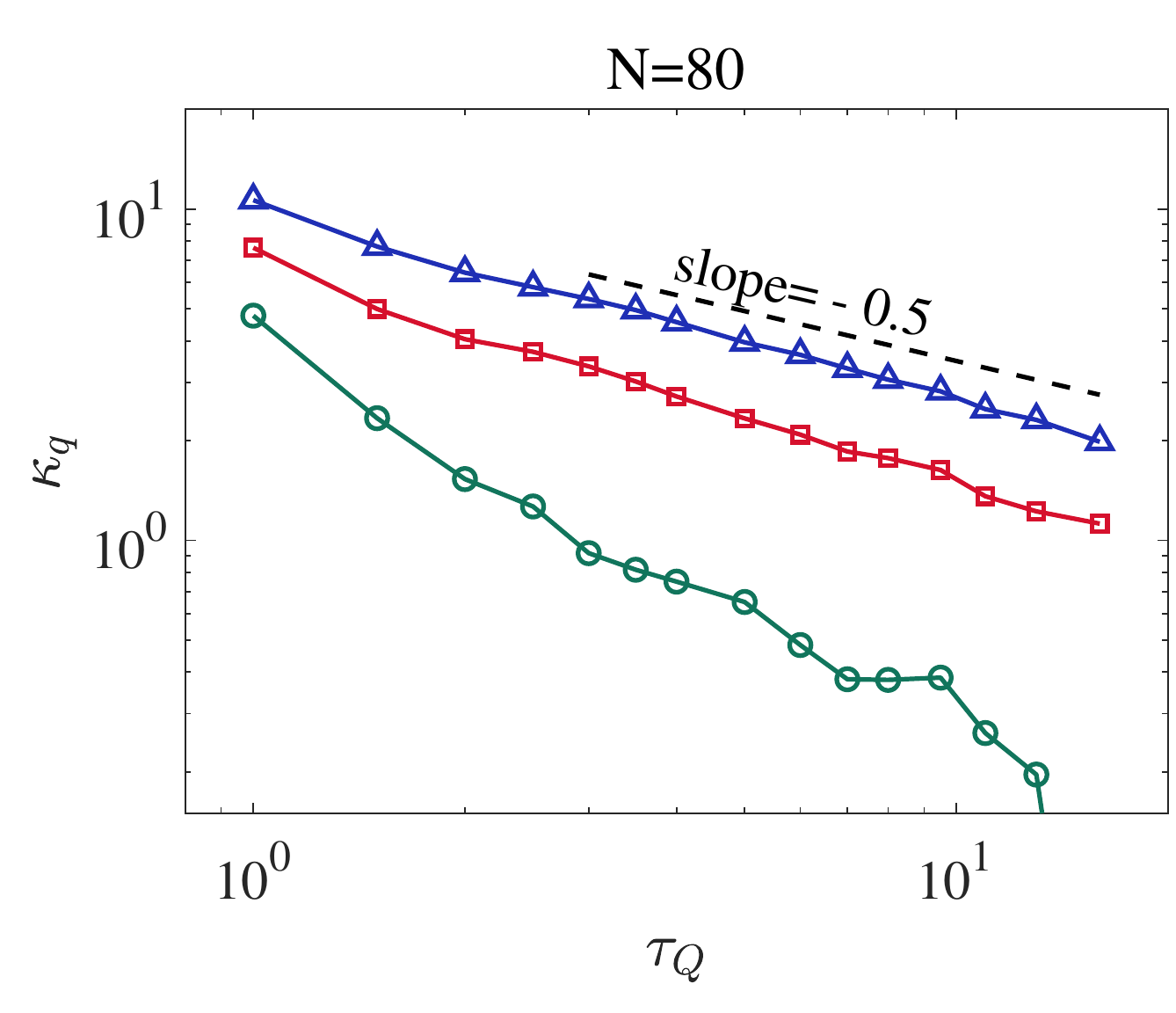}
	\includegraphics[trim=3cm 9cm 4.4cm 9cm, clip=true, scale=0.5, angle=0]{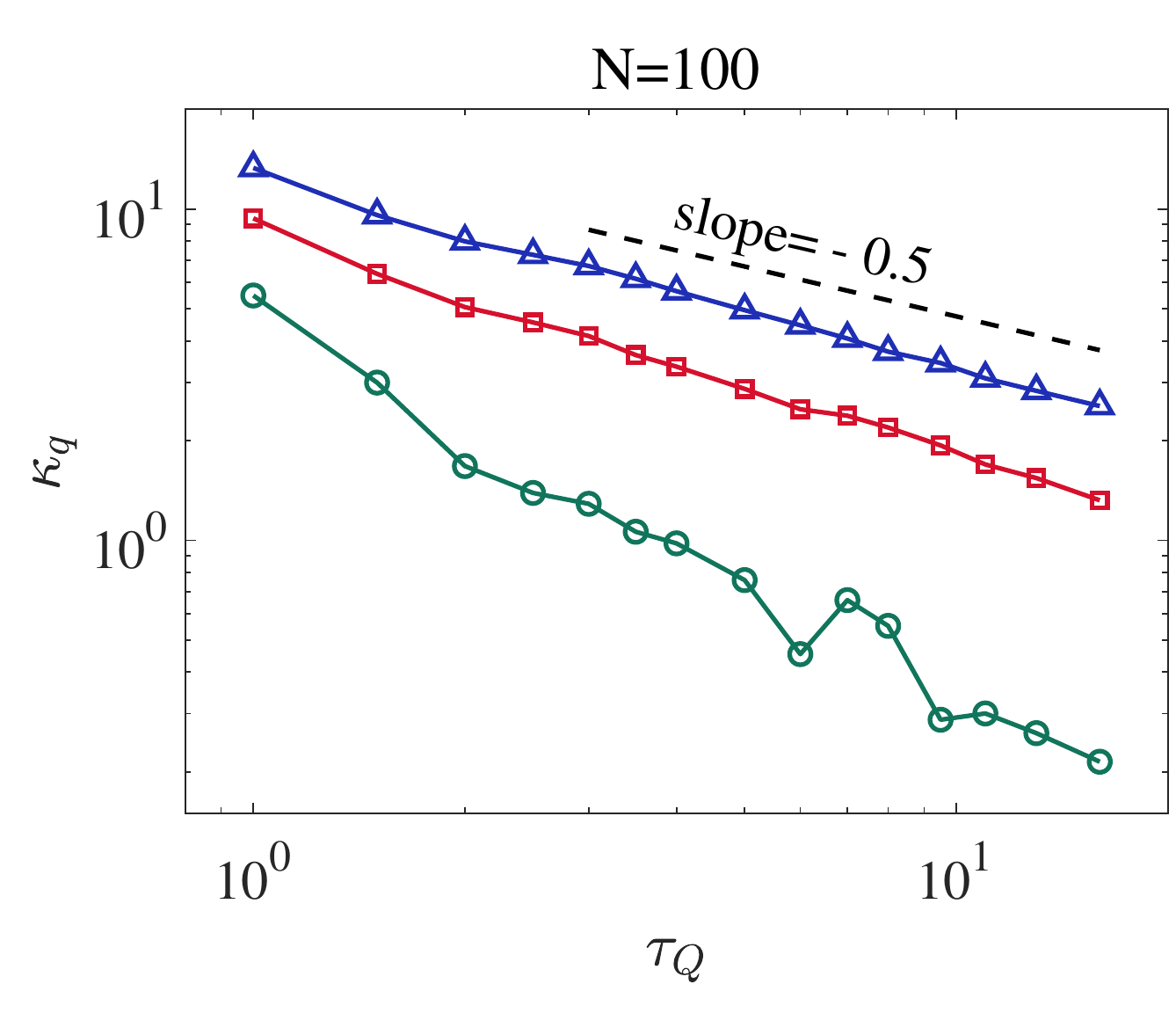}
	\caption{The first three cumulants against the quench rate for various sites numbers $N$. The dashed lines are the theoretical predictions with the power-law scaling $\tau_Q^{-0.5}$. The numerical data are from the neural network methods. }
	\label{FSE}
	\end{center}
\end{figure}

 In Fig.\ref{FSE}, we study the effects of $N$ on the first three cumulants $\kappa_{1,2,3}$ of the kink pair numbers. Specifically, we show four plots with $N=40, 60, 80$ and $100$, respectively. From the figure, we can see that as $N=40$, the first cumulant $\kappa_1$ is already very straight in the range $\tau_Q\geq3$, except the last point deviates a little bit away from the line. However, as $N$ grows we see that the last point of $\kappa_1$ will align in the straight line which is roughly parallel to the theoretical predictions $\tau_Q^{-0.5}$. For $\kappa_2$, we see that as $N=40$, $\kappa_2$ deviates from the straight line $\tau_Q^{-0.5}$ greatly as $\tau_Q$ is relatively big. However, for $N=60, 80$ and $100$, the behavior of $\kappa_2$ improves and tend to parallel with the theoretical prediction.  For $\kappa_3$, as $N=40$ it even  disappear for greater $\tau_Q$ since in this case $\kappa_3$ was computed to be negative. However, as $N$ increases we see that more and more data of $\kappa_3$ will appear and tend to align in the straight line $\tau_Q^{-0.5}$. But as we stated in the main text, $\kappa_3$ needs much more data to get a better behavior. $N=100$ is still not big enough for a perfect $\kappa_3$. However, as we have shown in Fig.\ref{FSE}, we can vividly see that as $N$ increases, the behavior of $\kappa_{1,2,3}$ will certainly improve and tend to be parallel with the line $\tau_Q^{-0.5}$. 
 
\begin{figure}[h]
	\begin{center}
		\includegraphics[trim=3cm 9cm 3cm 9cm, clip=true, scale=0.6, angle=0]{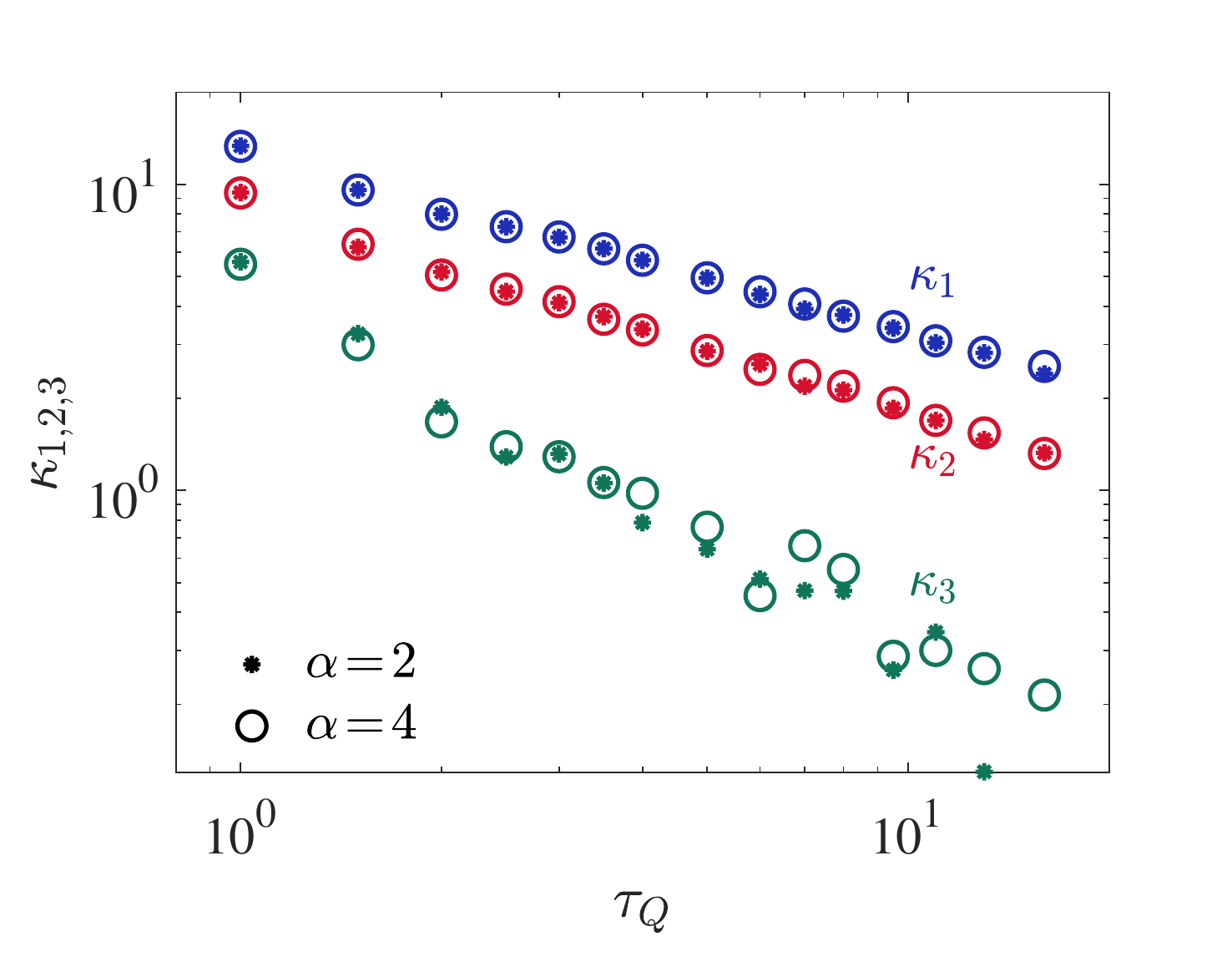}
		\caption{ The first three cumulants against the quench rate for various neural network sizes $\alpha$. The asterisks represent the cumulants for $\alpha=2$ while the circles for $\alpha=4$. We take $N=100$ in this figure. }
		\label{Error1}
	\end{center}
\end{figure}

\begin{figure*}[t]
	\begin{center}
		\includegraphics[trim=0cm 0cm 0cm 0cm, clip=true, scale=0.5, angle=0]{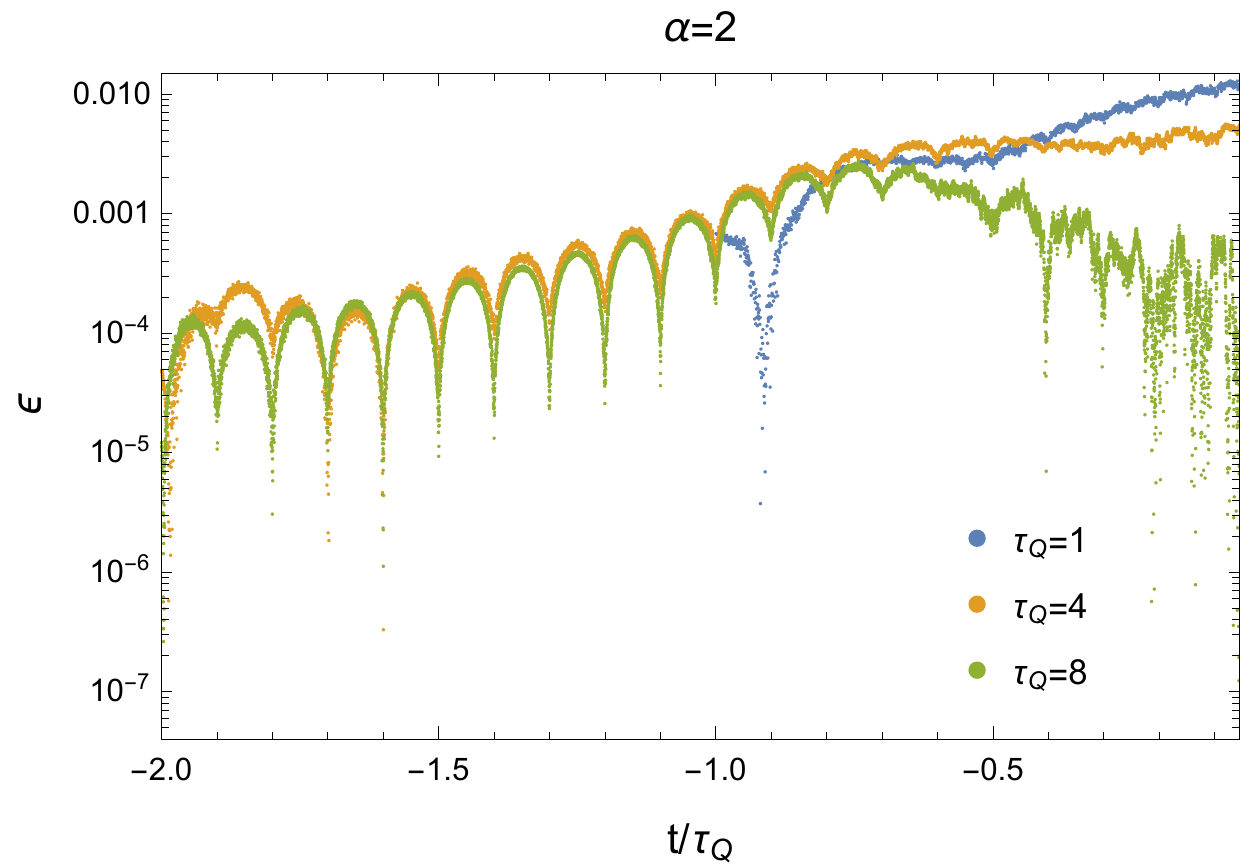}
		\includegraphics[trim=0cm 0cm 0cm 0cm, clip=true, scale=0.5, angle=0]{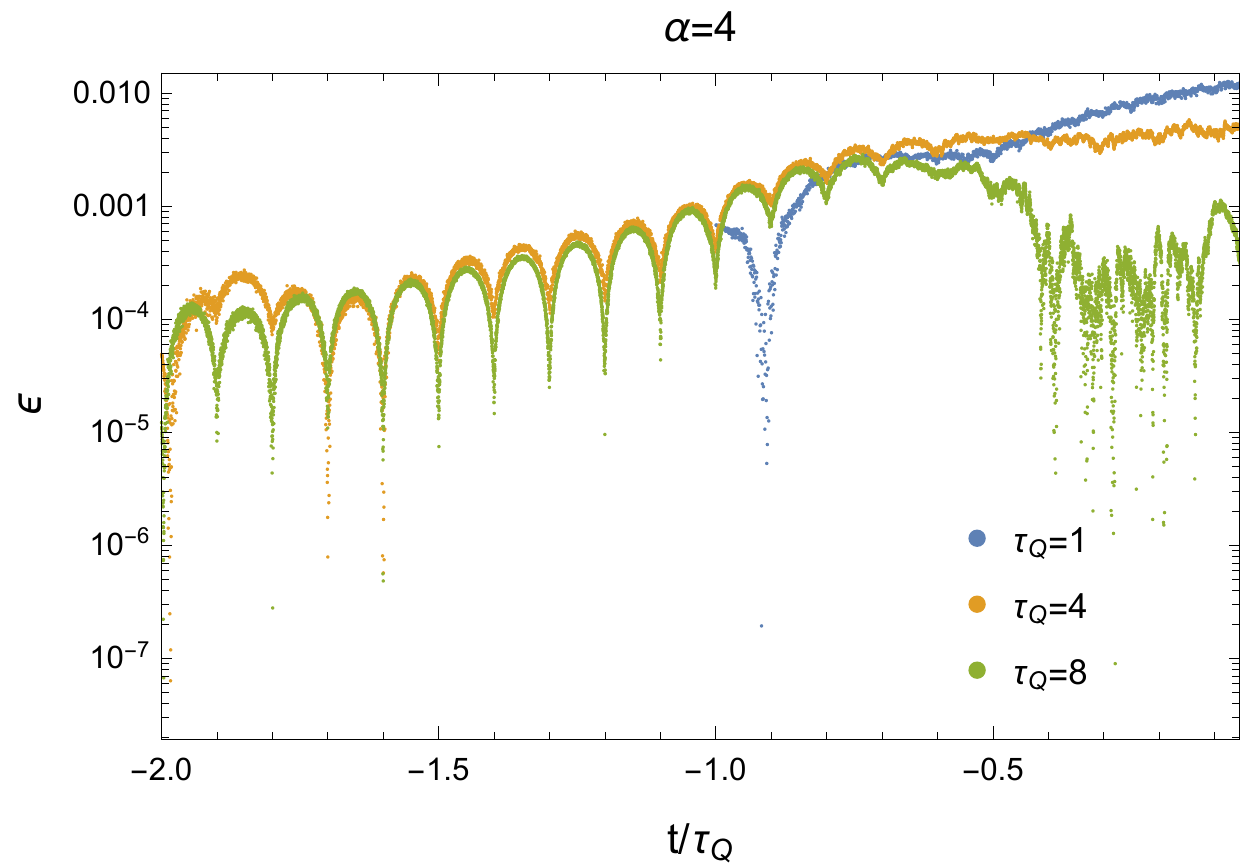}
		\caption{Time evolutions of the relative errors of the energies $\epsilon$ for $\alpha=2$ (left) and $\alpha=4$ (right).  }
		\label{Error2}
	\end{center}
\end{figure*}

\section{Effects of Neural Network Size on the Cumulants and the Energy Expectation Value}
\label{sec:b}
In this part, we will show the effects of the neural network size on the cumulants and the energy expectation value. We define the neural network size $\alpha$ as the ratio between the number of hidden neuron and the number of visible neuron, i.e, $\alpha=M/N$.  In Fig.\ref{Error1}, we show the effects of $\alpha$ on the cumulants. Specifically, we take $\alpha=2$ and $4$. From the Fig.\ref{Error1} we see that in the range of $\alpha\in[2,4]$, it has minor effects to the cumulants $\kappa_{1}$ and $\kappa_2$. However, $\alpha$ has strong effect on $\kappa_3$. It is clear that as $\alpha=4$, the behavior of $\kappa_3$ is much better than that of $\alpha=2$. Considering the cost of time and the resources of the computer, we take $\alpha=4$ in the main text.

Fig.\ref{Error2} shows the time evolutions of the relative errors of the energy for $\alpha=2$ and $4$. The relative error is defined as $\epsilon=(E_{\rm neural~network}-E_{\rm exact})/E_{\rm exact}$, where $E_{\rm exact}$ is from the analytic methods in \cite{dziarmaga2005dynamics}.  From Fig.\ref{Error2} we see that for $\alpha\in[2,4]$, there is no much distinctions between the relative error of the energy. Thus, from both Fig.\ref{Error1} and Fig.\ref{Error2}, it suggests that the neural network size $\alpha=4$ we take is precise enough for the TFQIM.


\end{document}